\documentclass[twocolumn,showpacs,showkeys,preprintnumbers,amsmath,amssymb,prl]{revtex4}
\usepackage{graphicx}
\usepackage{epsfig}

\begin{document}

\title{ Berry-curvatures and the anomalous Hall effect in Heusler compounds}

\author{J\"urgen K\"ubler$^1$ and Claudia Felser$^2$}

\affiliation{  $^1$Institut f\"{u}r Festk\"{o}rperphysik, Technische
Universit\"{a}t Darmstadt, 64289 Darmstadt, Germany,\\
  $^2$Institut f\"ur Anorganische Chemie und Analytische
  Chemie, Johannes Gutenberg - Universtit\"{a}t,  55099 Mainz}

\email{jkubler@fkp.tu-darmstadt.de}

\date{\today}

\pacs{75.10.Lp,75.47.Np,75.70.Tj}
\keywords{Berry curvature, Hall effect, Half-metallic ferromagnets, Heusler compounds}

\begin{abstract}
Berry curvatures are computed for a set of Heusler compounds using
density functional (DF)
 calculations and the wave functions that DF provide.
 The anomalous Hall conductivity is  obtained from the Berry curvatures.
 It is  compared with experimental values in the case of Co$_2$CrAl and Co$_2$MnAl.
A notable trend cannot be seen but the range of values is quite
enormous. The results for the anomalous Hall conductivities and
their large variations can be qualitatively understood by means of
the band structure and the Fermi-surface topology. \end{abstract}

\maketitle

A rather recent discovery is that the Berry-curvature supplies an
additional term to the electron velocity in a crystal \cite{xiao}.
It is an important correction to all transport properties that
rely on the velocity, in particular, it describes the leading
contribution to the anomalous Hall effect \cite{nagaosa}.

Heusler compounds, especially those based on Co, with their
regularities in many physical properties, like the Slater-Pauling
behavior \cite{kub}, invite the question if such regularities are
also present in the anomalous Hall effect (AHE). This is one of
the questions we turn to here. Furthermore, a considerable number
of Heusler compounds are half-metallic ferromagnets, $i.e.$ they
are gapped in one spin channel; therefore, in notable applications
one tries to make use of spin currents, for which our calculations
can serve as guide lines in estimates of the degree of spin
polarization of the current.

The Berry curvature follows from the Berry vector,
\begin{equation}\label{eq1}
\mathcal{A}(\mathbf{k})=i\sum_n \langle{u_{{\bf k},n}}|\nabla_k|{u_{{\bf k},n}}\rangle ,
\end{equation}
where $u_{{\bf k},n}(\mathbf{r})$ is the crystal-periodic
eigenfunction having wave vector \textbf{k} and band index $n$.
The sum extends over the occupied states which for metals vary
with \textbf{k}. The Berry curvature is written as
\begin{equation}\label{eq2}
\Omega(\mathbf{k})=\nabla_{\mathbf{k}}\times\mathcal{A}(\mathbf{k}).
\end{equation}
The Berry curvature can be calculated in different ways. The
common procedure is via a Kubo-like approach, where one calculates
essentially a Green's function, see e.g. ref. \cite{yao}. The
other, less common approach, is via the wave-functions directly.
The numerical treatment used here to calculate $\Omega$ is
basically a finite-difference approach, some details may be
summarized using the references \cite{king,resta,fukui,stocks} as
follows. One begins by computing the so-called link-variable
\cite{fukui}
\begin{equation}\label{eq2a}
U_{\mathbf{j}}(\mathbf{k})=\mathrm{det}[\langle u_{n \mathbf{k}}|u_{m \mathbf{k+j}}\rangle],
\end{equation}
where the determinant is evaluated for the occupied states $n$ and $m$.
The component of $\mathcal{A}(\mathbf{k})$ along $\mathbf{j}$ is then
\begin{equation}\label{eq2b}
\mathcal{A}_{\mathbf{j}}(\mathbf{k})=\mathrm{Im}\,
\mathrm{log}U_{\mathbf{j}}(\mathbf{k}),
\end{equation}
which yields by finite differences for the $z$-component of the
Berry curvature (except for a scaling factor)
\begin{equation}\label{eq2c}
\Omega_z(\mathbf{k})=\mathrm{Im}\,
\mathrm{log}\frac{U_y(\mathbf{k}+\mathbf{\hat{k}_x})U_x(\mathbf{k})}
{U_y(\mathbf{k})U_x(\mathbf{k}+\mathbf{\hat{k}_y})}.
\end{equation}
The logarithm implies that the results are mod $2\pi$.
\begin{figure}[h]
\begin{center}
\epsfig{file=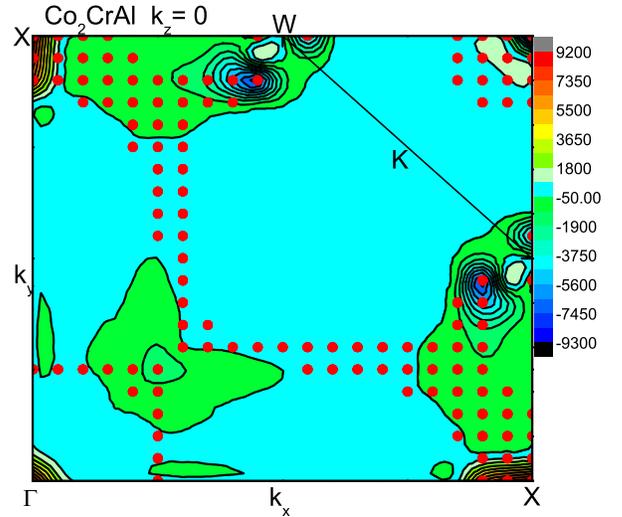,width=8cm} \caption{The Berry-curvature
in the $k_z$=0 plane for Co$_2$CrAl. The color codes are in units
of ($\Omega$cm)$^{-1}$. The labels follow the standard notation
for the face-centered-cubic crystal. The red dots mark band
energies of majority-spin electrons within 40 meV below the Fermi
energy.} \label{fig1}
\end{center}
\end{figure}

\begin{figure}[h]
\begin{center}
\epsfig{file=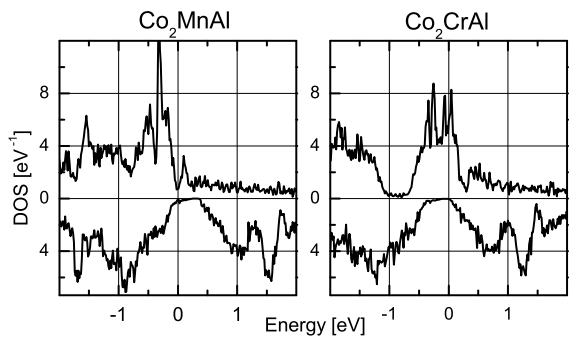,width=9cm} \caption{Density of states of
Co$_2$MnAl and Co$_2$CrAl. The upper parts of the figures describe
majority-spin electrons the lower parts minority-spin electrons.
In contrast to spin up and down, these terms are well defined even
in the spin-orbit coupled systems. } \label{fig5}
\end{center}
\end{figure}

The Heusler compounds of interest here are face-centered cubic
possessing the $L_{21}$ symmetry, except for Mn$_2$PtSn which is
tetragonal with space-group No. 119. They are ferromagnetic;
time-reversal symmetry is therefore broken which leads to a
non-zero Berry curvature \cite{xiao,nagaosa}.

The wave functions are calculated in the local density-functional
approximation (LDA) \cite{kohn} using the ASW method
\cite{williams}. Spin-orbit coupling (SOC), which is essential for
this theory, is included in a second variation \cite{mcdonald}.

The anomalous Hall conductivity, $\sigma_{xy}$, is given by the
Berry curvature as
\begin{equation}\label{eq3}
\sigma_{xy}=-\frac{e^2}{\hbar}
\,\frac{1}{N}\sum_{\mathbf{k}\in(BZ)}
\Omega_z(\mathbf{k})f(\mathbf{k}),
\end{equation}
where $f(\mathbf{k})$ is the Fermi distribution function,
$\Omega_z(\mathbf{k})$ is the $z$-component of the berry curvature
for the wave-vector \textbf{k}, $N$ is the number of electrons in
the crystal and the sum extends over the Brillouin zone (BZ).

In the figures to be presented the $z$-component of the
Berry-curvature is shown in a cut through the fcc Brillouin (BZ)
Zone; this was chosen to be the $k_z$=0 plane. While the number of
k-points (441) to obtain these figures could be chosen
sufficiently large for the plots to show significant details, the
Hall conductivity, $\sigma_{xy}$, at $T=0$ K which was calculated
by means of Eq. (\ref{eq3})  needed very large numbers of k-points
for convergence. Our results are converged to within about 20 \%
with about 2000 points in the irreducible wedge of the BZ.

Haldane \cite{haldane} showed that Fermi-liquid theory is still
valid even though it appears that the Hall conductivity, Eq.
(\ref{eq3}), depends on all states below the Fermi energy. He
showed that the Berry curvature can be transformed to the Berry
phase on the Fermi-surface only. Thus an alternative to our
calculations exists \cite{nagaosa}. This is a transformation to
Berry phases on the Fermi-surface, but it requires good software
to handle the three dimensional surfaces. Our calculations, in
contrast, are indeed quite straight forward. The results of our
calculations are collected in Table. \ref{t1}, where also an
estimate, $P$, of the degree of spin-polarization of the Hall
current is given. This is obtained by counting the number of
majority-spin electron states within 40 meV below the Fermi
energy, $N^+$, in the irreducible wedge of the BZ, and similarly
the number of minority-spin states, $N^-$, then
$P=N^{+}/(N^{+}+N^{-})$. These numbers are given by
$N^{\pm}=\sum_{\mathbf{k}}n^{\pm}(\mathbf{k})$, where the
spin-resolved norms, $n^{\pm}(\mathbf{k})$, are either 0 or 1, but
SOC mixes into a given spin state contributions of the opposite
spin, thus a "spin filter" finds the $n^{\pm}(\mathbf{k})$ larger
than 0 or smaller than 1.

\begin{table}[h]
\caption{\label{t1} Collection of experimental and calculated data
relevant for the Hall conductivity. $N_V$ is the number of valence
electrons, $M^{\rm exp}$ is the the experimental and $M^{\rm
calc}$ the calculated magnetic moment in $\mu_B$, $\sigma_{xy}$ is
the Hall conductivity in $(\Omega \mathrm{cm})^{-1}$, calculated
by means of eq.(\ref{eq3}), $P$ is an estimate of the spin
polarization of the Hall current.}
 \begin{center}
\begin{tabular}{lccccc}
\hline \hline \\
Compound & $N_V$& $M^{\rm exp}$ & $M^{\rm calc}$ & $\sigma_{xy}$ & $P$ (\%)  \\
\hline \\
Co$_2$VGa  & 26  &  1.92 & 1.953 & 66 & 65  \\
Co$_2$CrAl & 27  &  1.7  & 2.998 & 438 & 100 \\
Co$_2$VSn  & 27  &  1.21 & 1.778  & -1489 & 35 \\
Co$_2$MnAl & 28  &  4.04 & 4.045 & 1800 & 75 \\
Rh$_2$MnAl$^{(1)}$ & 28  &   -- & 4.066 & 1500 & 94 \\
Mn$_2$PtSn$^{(2)}$& 28  &   -- & 6.66 & 1108 & 91 \\
Co$_2$MnSn & 29  &  5.08 & 5.00  & 118 & 82 \\
Co$_2$MnSi & 29  &  4.90 & 4.98  & 228 & 100\\
\hline
\end{tabular}
\end{center}
\begin{center}
\begin{tabular}{lp{12cm}}
\noindent \footnotesize  \footnotesize $^{(1)}$  Crystal data from Pearson's Crystal Data.\\
\noindent \footnotesize  \footnotesize $^{(2)}$ Crystal data from
J. Winterlik (private communication).
 \\
\end{tabular}
\end{center}
\end{table}

\begin{figure}[h]
\begin{center}
\epsfig{file=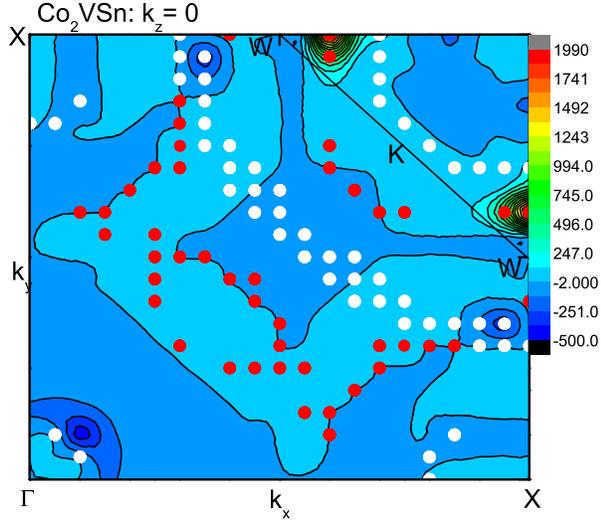,width=8cm} \caption{The Berry-curvature in
the $k_z$=0 plane for Co$_2$VSn. Color code and labels as in Fig.
\ref{fig1}. The red and white dots mark band energies of
majority-spin and minority-spin electrons, respectively, within 40
meV below the Fermi energy. } \label{fig2}
\end{center}
\end{figure}

\begin{figure}[h]
\begin{center}
\epsfig{file=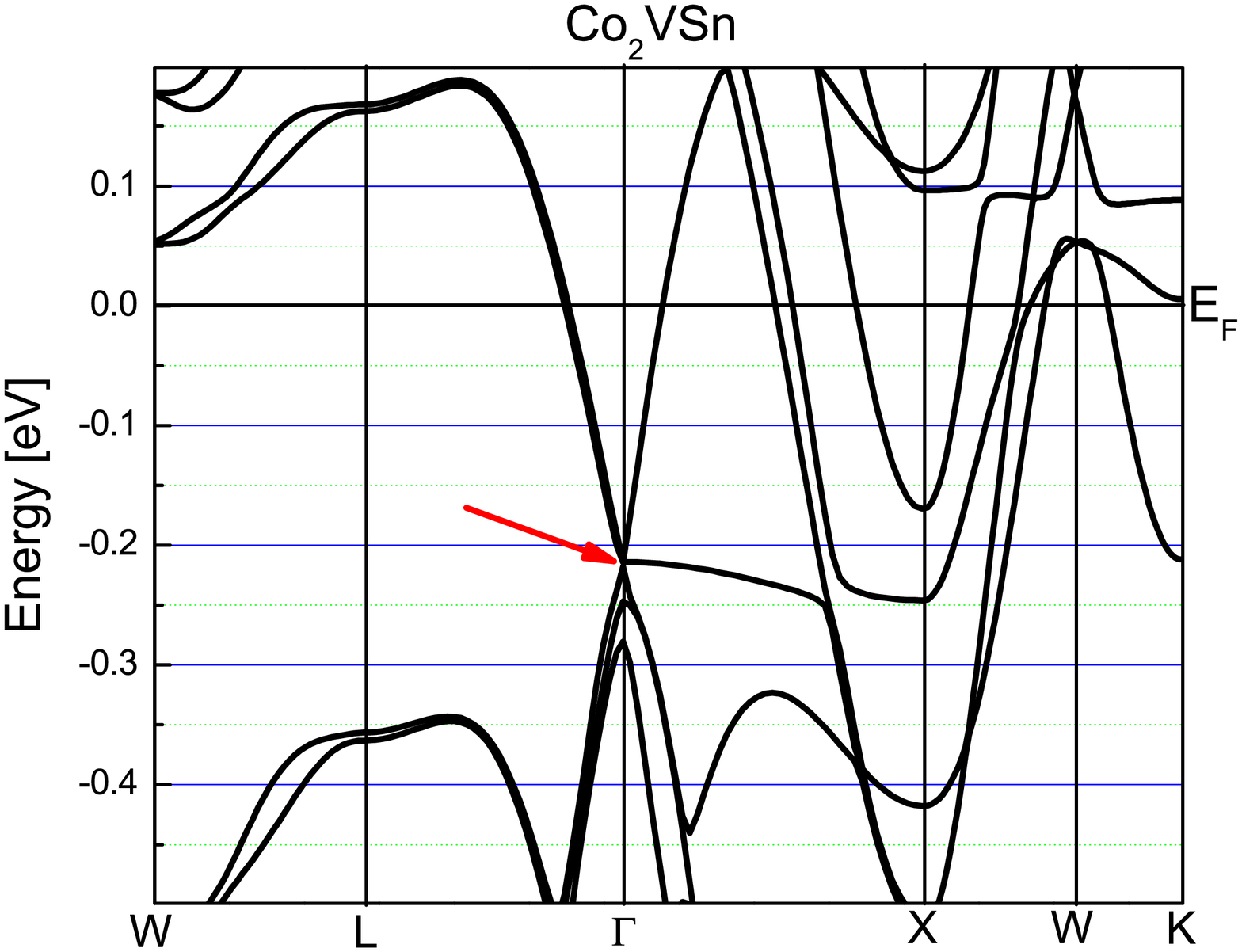,width=9cm} \caption{Band structure near
the Fermi edge of Co$_2$VSn. The red arrow points toward the Dirac
point. } \label{fig3}
\end{center}
\end{figure}

\begin{figure}[h]
\begin{center}
\epsfig{file=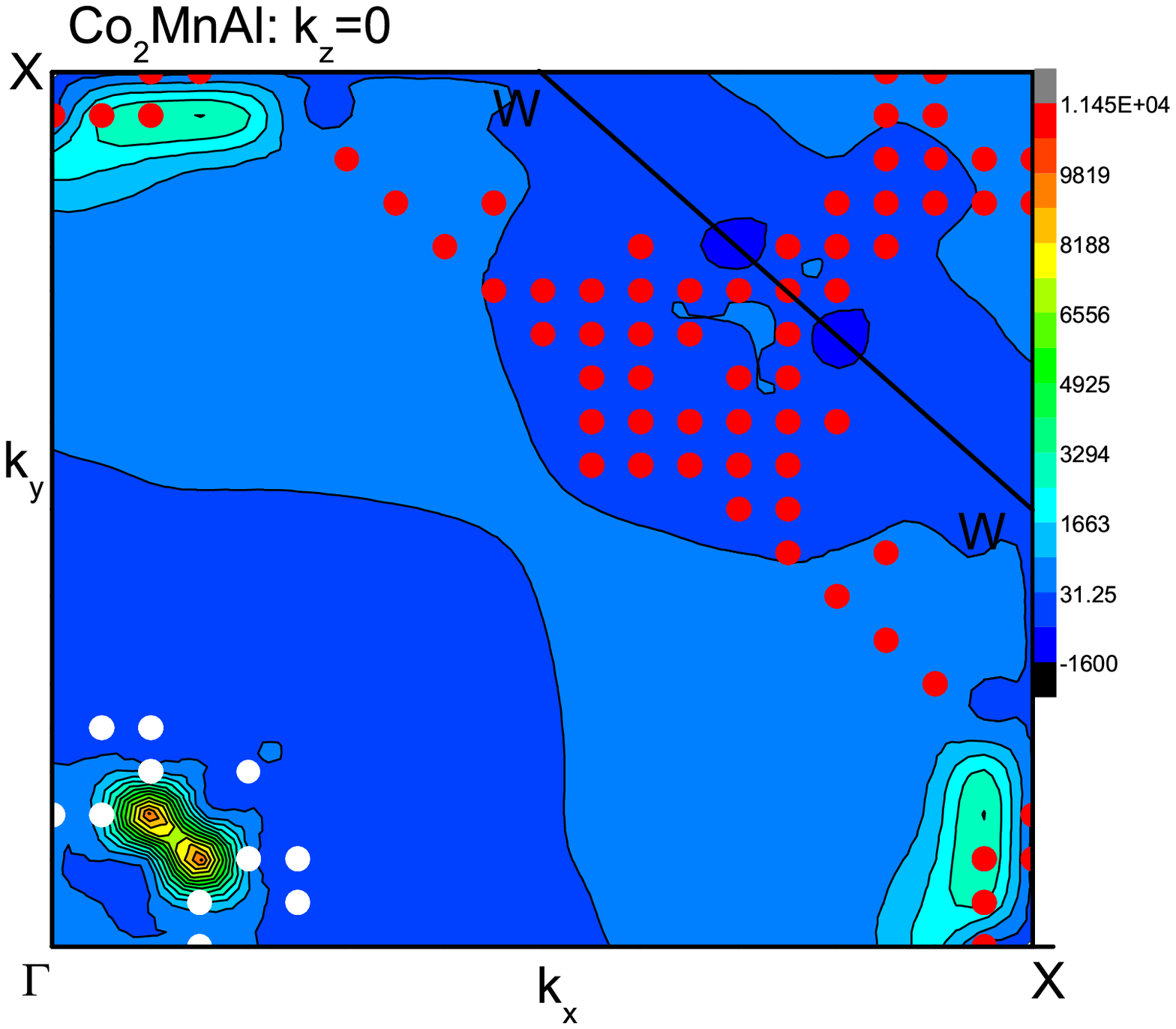,width=8cm} \caption{The Berry-curvature in
the $k_z$=0 plane for Co$_2$MnAl. Color code, labels and dots as
in Fig.\ref{fig2}. } \label{fig4}
\end{center}
\end{figure}

\begin{figure}[h]
\begin{center}
\epsfig{file=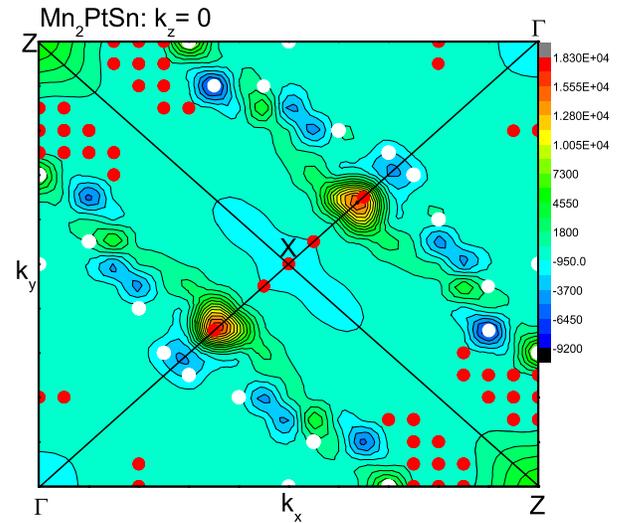,width=8cm} \caption{The Berry-curvature in
the $k_z$=0 plane for Mn$_2$PtSn. Color code and dots as in Fig.
\ref{fig2}. The symmetry labels are the standard ones for the bct
lattice. } \label{fig6}
\end{center}
\end{figure}

\begin{figure}[h]
\begin{center}
\epsfig{file=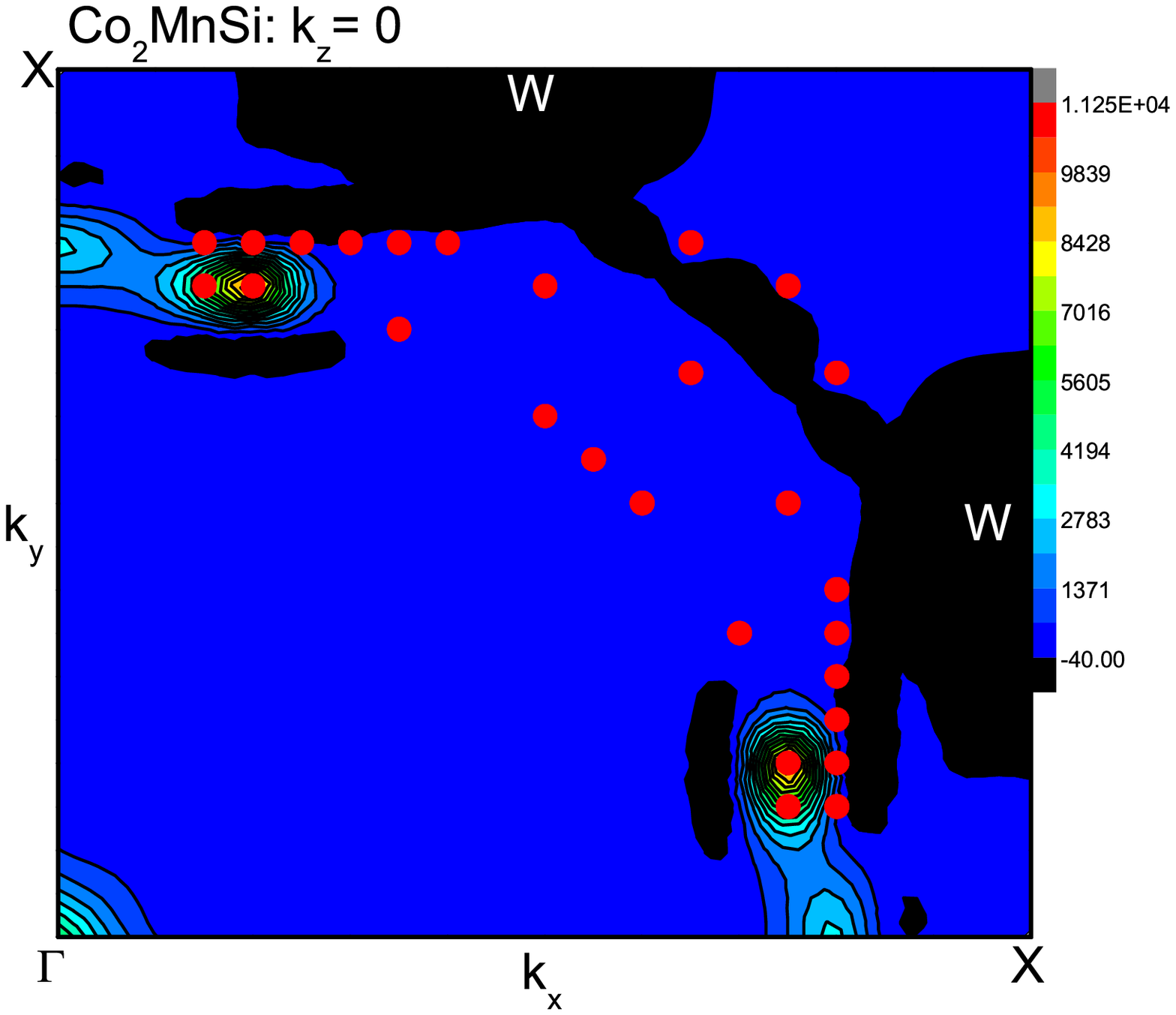,width=8cm} \caption{The Berry-curvature in
the $k_z$=0 plane for Co$_2$MnSi. Color code, labels and dots as
in Fig.\ref{fig2}.} \label{fig7}
\end{center}
\end{figure}

 Starting with the valence electron number $N_V$=26, the Hall
conductivity is calculated for Co$_2$VGa and is given in Table
\ref{t1}. The Berry curvature for Co$_2$CrAl with $N_V$=27 valence
electrons is shown in Fig.\ref{fig1} and the integrated value is
given in Table \ref{t1}. All states within 40 meV below the Fermi
energy are majority-spin states (red dots in the figure) which
agrees with the value of $P$ given in the Table. The density of
states of Co$_2$CrAl shown in Fig.\ref{fig5} also agrees with
$P=100 \%$. There is an experimental value of $\sigma_{xy}=125 \,\
(\Omega \mathrm{cm})^{-1} $ to be compared with our value of
$\sigma_{xy}=472 \,\ (\Omega \mathrm{cm})^{-1} $. The difference
is significant, but the rather low value measured for the magnetic
moment of 1.7 $\mu_B$ cannot be explained by our density of states
shown in fig. (\ref{fig5}) which results in a magnetic moment of 3
$\mu_B$. The reason is most likely that the sample is disordered
and does not have the ideal Heusler $L_{21}$ crystal structure.
Furthermore, it is possible but less likely that the other
contributions to the Hall effect, \textit{i.e.} side jump-and skew
scattering-mechanisms, contribute considerably.

The Heusler compound Co$_2$VSn also with $N_V$=27 valence
electrons is not a half-metallic ferromagnet \cite{kub} having a
measured moment of 1.21 $\mu_B$ and a calculated one of 1.78
$\mu_B$. In Fig.\ref{fig2} we show the Berry curvature in the
$k_{z}=0$ plane as before. The band-structure reveals in Fig.\ref{fig3} a Dirac-cone below the Fermi energy describing
minority-spin electrons. They show up in the berry curvature as
the semi circle and the white dots around the $\Gamma$-point. The
states seen near the $W$-points are due to majority-spin
electrons (red dots). The Dirac cone results in a large negative contribution
to the Hall conductivity, while the other states give less
positive contributions resulting in a calculated Hall conductivity
of $\sigma_{xy}=-1489\,\ (\Omega \mathrm{cm})^{-1}$ and a
polarization of only 35 \%.

Next is Fig.\ref{fig4} for Co$_2$MnAl with 28 valence electrons.
The magnetic moment is measured and calculated to be 4.04 $\mu_B$.
Here we obtain the startling high value of  $\sigma_{xy}=1800\,\
(\Omega \mathrm{cm})^{-1} $. In the density of states, Fig.
\ref{fig5}, the Fermi energy sits in tails of states at the
low-energy side of the gap, therefore we see down-spin electrons
in the Fermi surface near $\Gamma$ and up-spin electrons near $X$ (red dots).
Both contribute positive to $\sigma_{xy}$, while negative
contributions originate from states around $K$ and $W$. The Fermi
surface near $\Gamma$ shows up in Fig.\ref{fig4} as two small
imperfectly resolved circles and the white dots.

There is a recent study of the AHE for Co$_2$MnAl by Vidal
\textit{et al.} \cite{jakob} which allows a rough estimate of the
Hall conductivity. If one takes their measured saturation value of
the Hall resistivity of $\rho_{xy}=20 \mu \Omega \mathrm{cm}$ and
 their estimated specific resistivity of order of 100 $\mu \Omega
\mathrm{cm}$ then the Hall conductivity is obtained to be
approximately 2000 $(\Omega \mathrm{cm})^{-1}$. This could be
called to be in good agreement with the theory were it not for
experimentally disordered Mn and Al sites in the samples a fact
that is stressed by Vidal \textit{et al.} \cite{jakob}.

To guide the search for other ferromagnetic compounds with a large
Hall conductivity we enquired into the role of the strength of
SOC. For this reason we increased the SOC strength by 40\% and
found for Co$_2$MnAl a conductivity of $\sigma_{xy}=2150\,\
(\Omega \mathrm{cm})^{-1}$. This increase originates from the
states near $X$ in Fig.\ref{fig4} and from a diminishing
importance of the states near $K$. The increased role of SOC is
realized in Rh$_2$MnAl, for which, therefore, the Hall
conductivity was calculated and found to be 1500 $(\Omega
\mathrm{cm})^{-1}$. This is within the value obtained for
Co$_2$MnAl but does not show the expected increase.

Since it is the valence electron number of $N_V$ = 28 where the
conductivity is especially large, we calculated the electronic
structure of the tetragonal Heusler compound Mn$_2$PtSn which also has 28 valence electrons. The Berry
curvature is shown in Fig.\ref{fig6}, see also Table \ref{t1}. It
is seen that the minority-spin states result in negative
contributions to the Hall conductivity, the total
$\sigma_{xy}=1108\,\ (\Omega \mathrm{cm})^{-1}$ being only
reasonably large.

For Co$_2$MnSi, finally, the Berry curvature is shown Fig.
\ref{fig7}. This compound has 29 valence electrons, it is a
half-metallic ferromagnet having a measured magnetic moment of
4.90 $\mu_B$ which is calculated to be 4.98 $\mu_B$. The
spin-polarization of the Hall current is 100 \%, still, the Hall
conductivity is only $\sigma_{xy}=228\,\ (\Omega
\mathrm{cm})^{-1}$.

Summarizing we state that the strong trends like the
Slater-Pauling behavior are not seen in the Hall conductivity.
However, large values of the Hall conductivity can be found in
special cases like Co$_2$MnAl and other systems having 28 valence
electrons. The strength of SOC is shown to be but one ingredient
to ensure large values of $\sigma_{xy}$, however, a general rule
still has to be found. 
\begin{acknowledgments} The generous supply of computer time by U. Nowak (Universit\"at Konstanz) and the financial support by the
DFG/ASPIMATT project (unit 1.2-A) are gratefully acknowledged.
\end{acknowledgments}

\end{document}